\documentclass[twocolumn,showpacs,superscriptaddress]{revtex4}
\usepackage{amssymb}
\usepackage{amsfonts}
\usepackage{amsmath}
\usepackage{graphicx}

\begin{document}

\title{Non-Markovian entanglement dynamics of noisy continuous variable
quantum channels}
\author{Jun-Hong An}
\affiliation{Department of Physics and Center for Quantum
Information Science, National Cheng Kung University, Tainan 70101,
Taiwan } \affiliation{Department of Modern Physics, Lanzhou
University, Lanzhou 730000, People's Republic of China}
\author{Wei-Min Zhang}
\affiliation{Department of Physics and Center for Quantum Information Science, National
Cheng Kung University, Tainan 70101, Taiwan }
\affiliation{National Center for Theoretical Science, Tainan 70101, Taiwan}

\begin{abstract}
We investigate the entanglement dynamics of continuous-variable
quantum channels in terms of an entangled squeezed state of two
cavity fields in a general non-Markovian environment. Using the
Feynman-Vernon influence functional theory in the coherent-state
representation, we derive an exact master equation with
time-dependent coefficients reflecting the non-Markovian influence
of the environment. The influence of environments with different
spectral densities, e.g., Ohmic, sub-Ohmic, and super-Ohmic, is
numerically studied. The non-Markovian process shows its remarkable
influences on the entanglement dynamics due to the sensitive
time-dependence of the dissipation and noise functions within the
typical time scale of the environment. The Ohmic environment shows a
weak dissipation-noise effect on the entanglement dynamics, while
the sub-Ohmic and super-Ohmic environments induce much more severe
noise. In particular, the memory of the system interacting with the
environment contributes a strong decoherence effect to the
entanglement dynamics in the super-Ohmic case.
\end{abstract}

\maketitle

\section{Introduction}

Quantum teleportation incorporating the classical communication
theory with a unique characteristic of quantum mechanics, quantum
entanglement, has received tremendous attention in the study of
quantum communication in the past decade
\cite{Bennett93,Nielsen00,Braunstein05}. In quantum teleportation
protocols, a necessary ingredient is the quantum channel, which is
realized through an entangled quantum state of two systems separated
between the sender and the receiver. Theoretically, both the
discrete, (for example, two polarized photons, two-level atoms,
orthe spins of electrons
etc.)\cite{Bouwmeester97,Hagley97,Rowe01,Oliver02,Yamamoto03,Riebe04}
and the continuous-variable (coherent and squeezed optical fields)
\cite{Vaidman94,Braunstein98} entangled states are equally useful
for a quantum channel. Practically, compared with the
discrete-variable entangled state, the continuous-variable entangled
state may be more efficient because it has less decoherence
\cite{Furusawa98,Bowen03,Zhang03}. Continuous-variable entangled
states can be traced back to the original paper on quantum
entanglement by Einstein, Podolsky, and Rosen \cite{EPR35}, where
entangled states of the common eigenstate of relative position and
total momentum of two particles were proposed. Such ideal entangled
states can actually be realized by a two-mode squeezed state of
optical fields in the large limit of the squeezing parameter. In
fact, the entangled two-mode optical squeezed state has been
successfully produced via the nonlinear process of parametric down
conversion \cite{Ou92}. This triggered a variety of experiments
\cite{Furusawa98,Bowen03,Zhang03} applying such an entangled state
to quantum teleportation. The entangled two-mode optical squeezed
state has been of key importance as an entangled resource for
practical implementations of quantum-information protocols
\cite{Braunstein05}. However, a realistic analysis of any quantum
channel must take into account the noise effect from its
environment. There has been an increasing interest in describing
continuous-variable entanglement dynamics under noise
\cite{Jakub04,An05,Rossi06,Adesso,Mac05,Ban06,Maniscalco07,Chou07,An07,Goan07}.

The traditional approach to studying the environment-induced noise
effects treats the interaction between the quantum system and its
environment perturbatively, which yields approximate equations of
motion such as Redfield or master equations under the Born-Markov
approximation \cite{Redfield65,Lindblad76,Carmichael93}. Although
this treatment has been widely employed in the field of quantum
optics, where the characteristic time of the environmental
correlation function is much shorter compared with that of the
system investigated \cite{Carmichael93}, its validity is
experiencing more and more challenges in facing new experimental
evidences \cite{blatt07}. Moreover, the Born-Markov approximation is
in general invalid in dealing with most condensed-matter problems,
for example, a quantum system hosted in a nanostructured environment
\cite{John94,Aquino04,Budini06,mtlee06,Falci05}, because large
coupling constants and long correlation time scales of the
environment both require a non-perturbative description. Therefore a
nonperturbative description of the non-Markovian dynamics in open
quantum systems has attracted much attention over recent years
\cite{Breuer02}.

In fact, recently, non-Markovian processes have been extensively
studied in the entanglement dynamics of two continuous-variable
systems, such as two harmonic oscillators or two-mode
electromagnetic fields, interacting with bosonic environments
\cite{Ban06,Maniscalco07,Chou07,An07,Goan07}. In \cite{Ban06}, the
non-Markovian entanglement dynamic of two-mode Gaussian states is
studied based on the master equation derived perturbatively using
the projection operator method up to the second order with respect
to the system-reservoir coupling constant (which actually
corresponds to the Born approximation). In \cite{Maniscalco07} the
dynamics of two harmonic oscillators interacting with two
uncorrelated reservoirs was formulated based on the Hu-Paz-Zhang
master equation of quantum Brownian motion \cite{Hu92} but the
non-Markovian entanglement dynamics is analyzed up to the second
order of the system-reservoir coupling constant with Ohmic
reservoirs. More recently, an exact master equation for two coupled
harmonic oscillators linearly interacting with a common reservoir
has been derived using the Feynman-Vernon influence functional
theory \cite{Feynman63,Caldeira83,Hu92} where the decoherence and
disentangled dynamics of a bipartite displaced Gaussian states is
studied within the Markovian approximation\cite{Chou07}. In a very
recent paper \cite{Goan07}, the non-Markovian master equation of two
coupled harmonic oscillators interacting with either two independent
reservoirs or a common reservoir was derived perturbatively up to
the second order of the system-reservoir coupling constant (i.e.,
also in the Born approximation), and the entangled dynamics of
two-mode Gaussian states was analyzed with Ohmic reservoirs as well.
In our previous work \cite{An07}, we have derived the exact master
equation for two coupled cavity fields under the influence of vacuum
fluctuation using the Feynman-Vernon influence functional theory in
the coherent-state path integral formalism \cite{wzhang90}, and
studied the decoherence dynamics of the continuous-variable quantum
channel in terms of entangled two-mode Glauber coherent states with
Ohmic spectral density.

In the present work, we shall explore the non-Markovian influence of
the vacuum fluctuation on the continuous-variable quantum channel in
terms of an entangled two-mode squeezed state with different
spectral densities, i.e. the Ohmic, the sub-Ohmic, and the
super-Ohmic cases. To study the non-Markovian entanglement dynamics
of the squeezed-state quantum channel under the influence of the
vacuum fluctuation, we model the system as two cavity fields
coupling to a common bosonic environment in the at zero temperature.
We then use the Feynman-Vernon influence functional theory in the
coherent-state path integral formalism that we have provided in our
previous work \cite{An07} to study nonperturbatively the noise
effect on the entanglement dynamics of the squeezed states. As is
well-known the Feynman-Vernon influence functional theory enables us
to treat both of the back actions from the environment to the system
and the system to the environment self-consistently. The dissipation
and noise dynamics of the quantum channel, going beyond the
Born-Markov approximation, is then governed by an effective action
associated with the influence functional containing all the
influences of the environment on the system.

Utilizing this nonperturbative treatment, the resulting exact master
equation can be expressed in an operator form with time-dependent
coefficients describing the full dynamics of the back action between
the system and the environment. We thereby investigate the
non-Markovian entanglement dynamics of the quantum channel under the
influence of environments with different spectral densities, i.e.
Ohmic, sub-Ohmic, and super-Ohmic densities. The influence of the
environment induces a shifted frequency $\Omega(t)$ and a decay rate
$\Gamma (t)$ in each cavity mode, as well as a shifted coupling
strength $\Omega'(t)$ and a correlated decay rate $\Gamma'(t)$
between the two modes. The entanglement dynamics depends sensitively
on the different shifted coupling strength $\Omega'(t)$
(noise-induced entanglement oscillation), and the decay rates
$\Gamma (t)$ and $\Gamma ^{\prime }(t)$ (dissipation-induced
suppression of quantum entanglement) for different spectral
densities. We find that the Ohmic environment shows a weak
dissipation-noise effect, while the sub-Ohmic environment leads to
fast decoherence in the entanglement dynamics. The super-Ohmic
environment has the strongest memory effect, which heavily
suppresses the entanglement of the squeezed-state quantum channel.

The paper is organized as follows. In Sec. II, we introduce the
model describing the non-Markovian entanglement dynamics of the
continuous-variable quantum channel in terms of an entangled
squeezed state, and we shall also briefly review the master equation
we derived in \cite{An07}. In Sec. III, we use logarithmic
negativity as an entanglement measure of continuous-variable states
to discuss the entanglement dynamics of the squeezed state. The
numerical results of the entanglement dynamics are given in Sec. IV,
where we also analyze in detail the influences of the environment
with different spectral densities on the quantum channel. Finally, a
brief summary is made in Sec. VI.

\section{ The Hamiltonian and the exact non-Markovian master equation}

\subsection{The model Hamiltonian}

Our system consists of two coupled cavity fields subject to a
common environment. The Hamiltonian of the total system is given
by \cite{Carmichael93,Hackenbroich03}
\begin{equation}
H=H_{S}+H_{E}+H_{I},  \label{HM}
\end{equation}%
where
\begin{eqnarray*}
&&H_{S}=\hbar \omega _{1}a_{1}^{\dag }a_{1}+\hbar \omega _{2}a_{2}^{\dag
}a_{2}+\hbar \kappa (a_{1}^{\dag }a_{2}+a_{2}^{\dag }a_{1}), \\
&&H_{E}=\sum_{k}\hbar \omega _{k}b_{k}^{\dagger }b_{k}, \\
&&H_{I}=\sum_{l,k}\hbar (g_{lk}a_{l}^{\dag }b_{k}+g_{lk}^{\ast
}a_{l}b_{k}^{\dag }),
\end{eqnarray*}%
are the Hamiltonians of the two cavity fields, the environment, and
their interaction, respectively. The operators $a_{l}$ and
$a_{l}^{\dag }$ ($l=1,2$) are the corresponding annihilation and
creation operators of the $l$th cavity field with frequency $\omega
_{l}$, and $\kappa $ is a real coupling constant between the two
cavity fields, which can be realized by a beam splitter. The
environment is modeled, as usual, by a set of harmonic oscillators
described by the annihilation and creation operators $b_{k}$ and
$b_{k}^{\dag }(k=1,2,\cdots )$. The coupling constants between the
cavity fields and the environment are given by $g_{lk}$. In the
present work we shall consider the entangled squeezed state used in
the quantum teloprotation of continuous-variable states
\cite{Furusawa98} where the two cavity fields are identical, i.e.,
$\omega _{1}=\omega _{2}\equiv \omega _{0}$. We also assume that the
dominant dissipation and noise effects are induced by the vacuum
fluctuation so that the environment is at zero temperature and the
two cavity fields should interact homogeneously with the
environment, namely, $g_{1k}=g_{2k}\equiv g_{k}$.

In order to investigate the decoherence effect in the entanglement
dynamics induced by the environment, a specification of the spectral
density $J(\omega )$ of the environment is required. The spectral
density characterizing the coupling strength of the environment to
the cavity fields with respect to its frequencies is defined by
\begin{equation}
J(\omega )=\sum_{k}|g_{k}|^{2}\delta (\omega -\omega _{k}).
\end{equation}%
In the continuum limit the spectral density may have the form
\begin{equation}
J(\omega )=\eta \omega (\frac{\omega }{\omega _{c}})^{n-1}\exp (-\frac{
\omega }{\omega _{c}}),  \label{spectral}
\end{equation}%
where $\omega _{c}$ is a cutoff frequency, and $\eta $ a dimensionless
coupling constant. The environment is classified as Ohmic if $n=1$,
sub-Ohmic if $0<n<1$, and super-Ohmic if $n>1$ \cite{CaldeiraRMP}.

\subsection{The exact master equation}

The exact master equation describing decoherence dynamics of the two
cavity fields can be derived with the Feynman-Vernon influence
functional method \cite{Caldeira83,Anastopoulos} in the
coherent-state representation \cite{wzhang90}. The detailed
derivation can be found in \cite{An07}, and we give only a few key
steps here for completeness. Going from the quantum mechanical
equation $i\hbar
\partial \rho _{\mathrm{tot}}(t)/\partial t=[H,\rho _{\mathrm{tot}}(t)]$,
the reduced density matrix fully describing the dynamics of the two cavity
fields is obtained by integrating out completely the environmental degrees
of freedom,
\begin{eqnarray}
\rho (\boldsymbol{\bar{\alpha}}_{f},\boldsymbol{\alpha
}_{f}^{\prime };t) &=&\int d\mu (\boldsymbol{\alpha }_{i})d\mu
(\boldsymbol{\alpha }
_{i}^{\prime})J(\boldsymbol{\bar{\alpha}}_{f}, \boldsymbol{\alpha
} _{f}^{\prime
};t|\boldsymbol{\bar{\alpha}}_{i},\boldsymbol{\alpha }
_{i}^{\prime };0)  \notag \\
&&~~~~~~~~~\times \rho
(\boldsymbol{\bar{\alpha}}_{i},\boldsymbol{\alpha }_{i}^{\prime
};0),  \label{rout}
\end{eqnarray}
where the reduced density matrix is obtained from the total density
matrix tracing over the environmental degrees of freedom, $\rho
(\boldsymbol{\bar{\alpha}}_{f},\boldsymbol{\alpha } _{f}^{\prime };
t )=\int d\mu (\mathbf{z})\langle
\boldsymbol{\alpha}_{f},\mathbf{z}|\rho _{\mathrm{tot}}\left(
t\right) | \boldsymbol{\alpha }_{f}^{\prime },\mathbf{z} \rangle $,
the complex variables $\boldsymbol{\alpha}= (\alpha_1,\alpha_2)$ and
$\mathbf{z}=({z_1, z_2, \cdots })$ the corresponding eigenvalues of
the cavity-field operators $a_{1,2}$ and the environment operators
$b_{k} (k=1,2, \cdots)$, acting on the bosonic coherent-state $|
\boldsymbol{\alpha},\mathbf{z}\rangle $, and
$\bar{\boldsymbol{\alpha }}$ denotes the complex conjugate of
$\boldsymbol{\alpha }$. The propagating function
$J(\boldsymbol{\bar{\alpha}}_{f},\boldsymbol{\alpha }_{f}^{\prime
};t| \boldsymbol{\bar{\alpha}}_{i},\boldsymbol{\alpha }_{i}^{\prime
};0)$, has the form
\begin{eqnarray}
J(\boldsymbol{\bar{\alpha}}_{f},\boldsymbol{\alpha }_{f}^{\prime };t|
\boldsymbol{\bar{\alpha}}_{i},\boldsymbol{\alpha }_{i}^{\prime };0)=\int
D^{2}\boldsymbol{\alpha }D^{2}\boldsymbol{\alpha }^{\prime }\exp \{\frac{i}{
\hbar }(S_{S}[\boldsymbol{\bar{\alpha}},\boldsymbol{\alpha }]  \notag \\
-S_{S}^{\ast}[\boldsymbol{\bar{\alpha}}',\boldsymbol{\alpha}'])\}
\mathcal{F[}\boldsymbol{\bar{\alpha}},\boldsymbol{\alpha
, \bar{\alpha}}',\boldsymbol{\alpha}'] , \label{J}
\end{eqnarray}
where $S_{S}[\boldsymbol{\bar{\alpha}},\boldsymbol{\alpha }]$ is the action
of the two cavity fields,
\begin{eqnarray}
S_{S}[\boldsymbol{\bar{\alpha}},\boldsymbol{\alpha }%
]&=&\hbar \sum_{l\neq l'}\Big\{-i \bar{\alpha}_{lf}\alpha _{l}(t)
+\int_{0}^{t}d\tau \big[ i
\bar{\alpha}_{l}(\tau)\dot{\alpha}_{l}(\tau )
 \nonumber \\ && ~~~~~~ -  \Delta
_{l}\bar{\alpha}_{l}(\tau)\alpha _{l}(\tau) - \kappa
\bar{\alpha}_{l}(\tau)\alpha _{l'}(\tau)\big]\Big\}, \nonumber
\end{eqnarray}
and $\mathcal{F[}\boldsymbol{\bar{\alpha}}, \boldsymbol{\alpha
,\bar{\alpha}}^{\prime },\boldsymbol{\alpha }^{\prime }]$ is the
Feynman-Vernon influence functional obtained after integrating out
all the degrees of freedom of the environment,
\begin{eqnarray}
\mathcal{F[}\boldsymbol{\bar{\alpha}},\boldsymbol{\alpha
,\bar{\alpha}}^{\prime },\boldsymbol{\alpha }^{\prime }]=\exp
\Big\{\int_{0}^{t}d\tau \int_{0}^{\tau }d\tau ^{\prime
}\Big[\sum_{l,m=1}^{2}(\bar{\alpha}'_{l}-\bar{\alpha}_{l})(\tau )  \notag \\
\times \mu(\tau -\tau')\alpha _{m}(\tau')+ (\alpha
_{l}-\alpha'_{l})(\tau )\mu^{\ast }(\tau -\tau')
\bar{\alpha}'_{m}(\tau') \Big]\Big\}. \notag
\end{eqnarray}
The time-dependent function $\mu(\tau)$ is the dissipation-noise
kernel characterizing the full influence of the environment on the
two cavity fields,
\begin{eqnarray}
\mu(\tau)=\sum_{k}e^{-i\omega _{k}(\tau)}|g_k|^2 = \int d\omega
J(\omega)e^{-i\omega (\tau)} ,
\end{eqnarray}
and is completely determined by the spectral density $J(\omega)$. We
only shall consider the spectral density given by (\ref{spectral})
in this paper.

As we see, all the effects of the environment on the system are
incorporated into the influence functional, which effectively
modifies the action of the cavity system. Since the resulting
effective action is bilinear in terms of the cavity field variables
$\boldsymbol{\alpha }$ and $\boldsymbol{\alpha }'$, the evaluation
of the path integral over $\boldsymbol{\alpha }$ and
$\boldsymbol{\alpha }'$ can be exactly executed with the saddle
point method. This leads to the dissipation-noise equations ($l\neq
l^{\prime } $ )
\begin{eqnarray}
\dot{\alpha}_{l}+i(\omega _{l}\alpha _{l}+\kappa \alpha _{l^{\prime }})
&=&-\int_{0}^{\tau }d\tau ^{\prime }\sum_{m=1}^{2}\mu \left( \tau -\tau
^{\prime }\right) \alpha _{m}\left( \tau ^{\prime }\right) ,  \notag \\
\dot{\bar{\alpha}}_{l}^{\prime }-i(\omega
_{l}\bar{\alpha}_{l}^{\prime }+\kappa \bar{\alpha}_{l^{\prime
}}^{\prime }) &=&-\int_{0}^{\tau }d\tau ^{\prime
}\sum_{m=1}^{2}\mu ^{\ast }\left( \tau -\tau ^{\prime }\right)
\bar{\alpha}_{m}^{\prime }\left( \tau ^{\prime }\right) ,  \notag \\
&&  \label{EOM}
\end{eqnarray}
obeying the boundary conditions $\alpha _{l}\left( 0\right)
=\alpha _{li}$ and $\bar{\alpha}_{l}^{\prime }\left( 0\right)
=\bar{\alpha}_{li}^{\prime }$. The integro-differential
dissipation-noise equations render the reduced dynamics
non-Markovian, with the memory of the system interacting with the
environment
 registered in the dissipation-noise kernel
$\mu(\tau-\tau^{\prime })$.
%The solution of the dissipation equations can be solved using
%Laplace transform and convolution theorem:
Introducing the new variables $u(t)$ and $v(t)$ by
\begin{eqnarray}
\alpha _{l}\left( \tau \right) &=&\alpha _{li}u\left( \tau \right) -\alpha
_{l^{\prime }i}v\left( \tau \right) ,\text{ \ \ }  \notag \\
\text{\ }\bar{\alpha}_{l}^{\prime }\left( \tau \right)
&=&\bar{\alpha} _{li}^{\prime }\bar{u}\left( \tau \right)
-\bar{\alpha}_{l^{\prime }i}^{\prime }\bar{v}\left( \tau \right)
,\text{\ \ \ }l\neq l^{\prime }, \label{soluen}
\end{eqnarray}%
we obtain an explicit solution for the propagating function,
\begin{eqnarray}  \label{prord}
&&J(\boldsymbol{\bar{\alpha}}_{f},\boldsymbol{\alpha }_{f}^{\prime };t|
\boldsymbol{\bar{\alpha}}_{i},\boldsymbol{\alpha }_{i}^{\prime };0)=  \notag
\\
&&~~~ \exp \Big\{\sum_{l=1}^{2}\big[u\bar{\alpha}_{lf}\alpha
_{li}+\bar{u} \bar{\alpha} _{li}^{\prime }\alpha _{lf}^{\prime }
-(\bar{u}u+\bar{v}v-1)
\bar{\alpha}_{li}^{\prime }\alpha _{li}\big]  \notag \\
&&~~~~~~~~-\sum_{l\neq l^{\prime }}\big[v\bar{\alpha}_{lf}\alpha
_{l^{\prime }i} +\bar{v}\bar{\alpha}_{li}^{\prime }\alpha
_{l^{\prime }f}^{\prime }-(
\bar{u}v+\bar{v}u)\bar{\alpha}_{li}^{\prime }\alpha _{l^{\prime
}i}\big]
\Big\}.  \notag \\
\end{eqnarray}

The non-Markovian master equation can be deduced from (\ref{rout})
and (\ref{prord}). The result is \cite{An07}
\begin{eqnarray}
\dot{\rho}(t) &=&-\frac{i}{\hbar }[H^{\prime }(t),\rho (t)]  \notag \\
&&+\Gamma (t)[2a_{1}\rho (t)a_{1}^{\dag }-a_{1}^{\dag }a_{1}\rho (t)-\rho
(t)a_{1}^{\dag }a_{1}]  \notag \\
&&+\Gamma (t)[2a_{2}\rho (t)a_{2}^{\dag }-a_{2}^{\dag }a_{2}\rho (t)-\rho
(t)a_{2}^{\dag }a_{2}]  \notag \\
&&+\Gamma ^{\prime }(t)[2a_{1}\rho (t)a_{2}^{\dag }-a_{1}^{\dag }a_{2}\rho
(t)-\rho (t)a_{1}^{\dag }a_{2}]  \notag \\
&&+\Gamma ^{\prime }(t)[2a_{2}\rho (t)a_{1}^{\dag }-a_{2}^{\dag }a_{1}\rho
(t)-\rho (t)a_{2}^{\dag }a_{1}],  \label{mas}
\end{eqnarray}%
where
\begin{equation*}
H^{\prime }(t)=\hbar \Omega (t)(a_{1}^{\dag }a_{1}+a_{2}^{\dag }a_{2})+\hbar
\Omega ^{\prime }(t)(a_{1}^{\dag }a_{2}+a_{2}^{\dag }a_{1}),
\end{equation*}%
with%
\begin{equation}
\frac{u\dot{u}-v\dot{v}}{u^{2}-v^{2}}\equiv -\Gamma
(t)-\frac{i}{\hbar } \Omega
(t),\frac{v\dot{u}-u\dot{v}}{u^{2}-v^{2}}\equiv -\Gamma ^{\prime
}(t)- \frac{i}{\hbar }\Omega ^{\prime }(t).  \label{go}
\end{equation}%
This is the exact master equation for the dynamics of the two cavity
fields, $\Omega (t)$ plays the role of a shifted time-dependent
frequency for each cavity field, $\Omega ^{\prime }(t)$ accounts for
a shifted time-dependent coherent coupling between the two cavity
fields, $\Gamma (t)$ represents a time-dependent individual decay
rate of each cavity field, and $\Gamma ^{\prime }(t)$ is a
correlated decay rate between the two cavity fields. From
Eq.~(\ref{mas}), we can see that besides the spontaneous decay of
the individual cavity field, the environment, even if only the
vacuum fluctuation is concerned, also induces a coherent coupling
and a correlated spontaneous decay between the two cavity fields.
The non-Markovian character thus resides in these time-dependent
coefficients in the master equation. We must emphasize that our
derivation of the master equation, Eq.~(\ref{mas}), is fully
nonperturbative, which goes beyond the Born approximation
\cite{Ban06,Goan07} and involves all the back actions between the
environment and the cavity fields.

\section{Entanglement measure of continuous-variable quantum channels and
its dynamics}

\subsection{Logarithmic negativity as entanglement measure}

In what follows, we shall analyze the effects of the different types
of noise on the entanglement dynamics of the quantum channel in
terms of an entangled two-mode squeezed state. The entangled
two-mode squeezed state is defined as the vacuum state acted on by
the two-mode squeezing operator
\begin{equation}
|\psi (0)\rangle =e^{r(a_{1}a_{2}-a_{1}^{\dag }a_{2}^{\dag })}|00\rangle ,
\label{initial}
\end{equation}
where $r$ is the squeezing parameter. The state approaches the ideal
Einstein-Podolsky-Rosen (EPR) state in the limit of infinite
squeezing ($r\rightarrow \infty $) \cite{EPR35}. After generating
the entangled state given by Eq.~(\ref{initial}), the two cavity
fields are then propagated, respectively, to the two locations
separated between the sender and the receiver. The quantum channel
is thus established through the entangled two-mode squeezed state
and is ready for teleporting unknown optical coherent states
\cite{Braunstein98,Furusawa98}. The traditional way to generate the
entangled two-mode squeezed state is via the nonlinear optical
process of parametric down-conversion \cite{Ou92}. Recently, a
microwave cavity QED-based scheme to generate such states has also
been proposed \cite{Pielawa07}.

To investigate the entanglement dynamics of the quantum channel in
terms of the two-mode squeezed state, a computable entanglement
measure for such continuous-variable states must be defined first.
Here we shall use the logarithmic negativity \cite{Werner02} to
quantify the degree of entanglement in the quantum channel. The
logarithmic negativity of a bipartite system was introduced
originally as
\begin{equation}
E_{N}=\log _{2}\sum_{i}\left\vert \lambda _{i}^{-}\right\vert ,
\end{equation}%
where $\lambda _{i}^{-}$ is the negative eigenvalue of $\rho
^{T_{i}}$, and $\rho ^{T_{i}}$ is a partial transpose of the
bipartite state $\rho $ with respect to the degrees of freedom of
the $i$th party. This measure is based on the Peres-Horodecki
criterion \cite{Peres96,Horodecki96} that a bipartite quantum state
is separable if and only if its partially transposed state is still
positive.

For the continuous-variable (Gaussian-type) bipartite state, its
density matrix is characterized by the covariance matrix defined as
the second moments of the quadrature vector
$X=(x_{1},p_{1},x_{2},p_{2})$,
\begin{equation}
V_{ij}=\frac{\langle \Delta X_{i}\Delta X_{j}+\Delta X_{j}\Delta
X_{i}\rangle }{2},
\end{equation}
where $\Delta X_{i}=X_{i}-\langle X_{i}\rangle $, and
$x_{i}=\frac{ a_{i}+a_{i}^{\dag }}{\sqrt{2}}$,
$p_{i}=\frac{a_{i}-a_{i}^{\dag } }{i\sqrt{2} }$. The canonical
commutation relations take the form as $ [X_{i},X_{j}]=iU_{ij}$,
with $\ U=\left( \begin{array}{cc} J & 0 \\ 0 & J \end{array}
\right) $ and $J=\left( \begin{array}{cc} 0 & 1 \\ -1 & 0
\end{array} \right) $ defining the symplectic structure of the
system. The property of the covariance matrix $V$ is fully
determined by its symplectic spectrum $ \nu =(\nu _{1},\nu _{2})$,
with $\pm \nu _{i}$ ($\nu _{i}>0$) the eigenvalues of the matrix:
$iU V$. The uncertainty principle exerts a constraint on $\nu _{i}$
such that $\nu _{i}\geqslant \frac{1}{2}$ \cite{Adesso05}. Thus the
Peres-Horodecki criterion for the continuous-variable state can be
rephrased as the state being separable if and only if the
uncertainty principle, $V+\frac{i}{2}U\geqslant 0$, is still obeyed
by the covariance matrix under the partial transposition with
respect to the degrees of freedom of a specific subsystem
\cite{Simon00}. In terms of phase space, the action of partial
transposition amounts to a mirror reflection with respect to one of
the canonical variables of the related subsystem. For instance,
$\tilde{V} =\Lambda V\Lambda $ , and $\Lambda =diag(1,1,1,-1)$ is
the partial transposition with respect to the second subsystem. If a
Gaussian-type bipartite state is nonseparable, the covariance matrix
$\tilde{V}$ will violate the uncertainty principle and its
symplectic spectrum $\tilde{ \nu}=(
\tilde{\nu}_{1},\tilde{\nu}_{2})$ will fail to satisfy the
constraint $ \tilde{\nu}_{i}\geqslant \frac{1}{2}$. The logarithmic
negativity is then used to quantify this violation as
\cite{Werner02}
\begin{equation}
E_{N}=\max \{0,-\log _{2}(2\tilde{\nu}_{\min })\},  \label{measu}
\end{equation}%
where $\tilde{\nu}_{\min }$ is the smaller one of the two symplectic
eigenvalues $\tilde{\nu}_{i}$. It is evident from Eq. (\ref{measu})
that, if $\tilde{ V}$ obeys the uncertainty principle, i.e.,
$\tilde{\nu}_{i}\geqslant \frac{1}{ 2}$, then $E_{N}(\rho )=0$,
namely, the state is separable. Otherwise, it is entangled.
Therefore, the symplectic eigenvalue $ \tilde{\nu}_{\min }$ encodes
a qualitative feature of the entanglement for an arbitrary
continuous-variable bipartite state.

\subsection{The entanglement dynamics}

With this entanglement measure at hand, we study now the
entanglement dynamics of the squeezed-state quantum channel in our
model. A straightforward way to obtain the time-dependent solution
of the entangled squeezed state is by integrating the propagator
function over the initial state of Eq. (\ref{rout}), where the
initial state in coherent-state representation is given by
\begin{equation}
\rho (\boldsymbol{\bar{\alpha}}_{i},\boldsymbol{\alpha }_{i}^{\prime };0)=
\frac{\exp [-\tanh r(\bar{\alpha}_{1i}\bar{\alpha}_{2i}+\alpha _{1i}^{\prime
}\alpha _{2i}^{\prime })]}{\cosh ^{2}r}.
\end{equation}%
The solution of the reduced density matrix can be obtained exactly,
\begin{eqnarray}
&&\rho (\boldsymbol{\bar{\alpha}}_{f},\boldsymbol{\alpha }_{f}^{\prime
};t)=b_{0}\exp [\sum_{l\neq l^{\prime }}(b_{1}\bar{\alpha}%
_{lf}^{2}+b_{2}\alpha _{lf}^{\prime 2}+b_{4}\bar{\alpha}_{lf}\alpha
_{lf}^{\prime }  \notag \\
&&~~~~~~+b_{5}\bar{\alpha}_{lf}\alpha _{l^{\prime }f}^{\prime }+\frac{b_{3}}{
2}\bar{\alpha}_{lf}\bar{\alpha}_{l^{\prime }f}+\frac{b_{6}}{2}\alpha
_{lf}^{\prime }\alpha _{l^{\prime }f}^{\prime })],  \label{final}
\end{eqnarray}%
where the time-dependent parameters $b_{i}$ ($i=0,\cdots ,6$) are
given explicitly in the Appendix.

From the above solution, the covariance matrix $V$ can be calculated
analytically, and the logarithmic negativity $E_{N}(t)$ can also be
obtained exactly from Eq. (\ref{measu}). It is easy to verify that
the initial entanglement is $E_{N}(0)=\frac{2r}{\ln 2}$. While the
asymptotical behavior of the entanglement (in the long time limit)
can be found from the solution
\begin{eqnarray}
\rho (t \rightarrow \infty ) &=&|\psi_{asy.} \rangle \langle \psi_{asy.} |,
\notag \\
|\psi_{asy.} \rangle &=&\sqrt{b^{\prime }_{0}}e^{b^{\prime }_{1}
(a_{1}^{\dag }-a_{2}^{\dag })^{2}}|00\rangle ,  \label{steady}
\end{eqnarray}%
where $b^{\prime }_{0}=b_{0}(t\rightarrow \infty )=\frac{1}{\cosh
r}$ and $ b^{\prime }_{1}= b_{1}(t\rightarrow \infty )=\frac{\tanh
re^{-2i(\omega _{0}-\kappa )t}}{4}$. This asymptotical solution
results in
\begin{equation}
E_{N}(t\rightarrow \infty )=\frac{r}{\ln 2},  \label{dfe}
\end{equation}
namely, the final entanglement is only one half of the initial
value. Equation (\ref{dfe}) indicates that the environment does
decrease the entanglement of the quantum channel, as a
dissipation-noise effect. This asymptotical result is also
consistent with the Markovian limit at zero temperature in
\cite{Jakub04}, since the non-Markovian dynamics must be
asymptotically reduced to the Markovian limit \cite{An07}.

It should be noted that the noise behavior of the quantum channel
also depends on the structure of the initial two-mode squeezed
state. This can be easily seen if we introduce the operators related
to the center-of-mass and relative motional variables of the two
cavity fields as $A^{\dag }=(a_{1}^{\dag }+a_{2}^{\dag })$ and
$a^{\dag }=(a_{1}^{\dag }-a_{2}^{\dag }) $, respectively. Then the
initial state can be rewritten in terms of these two operators,
\begin{equation}
|\psi (0)\rangle ={\frac{1 }{\cosh r}}e^{-\frac{\tanh r}{4}(A^{\dag
2}-a^{\dag 2})}|00\rangle .
\end{equation}
Since the two cavity fields interact homogeneously with the
environment, namely, $g_{1k}=g_{2k}=g_{k}$, the interaction between
the cavity fields and the environment only influences the dynamics
of the center-of-mass variable; it has no effect on the relative
motion of the two cavity fields (a similar discussion for two
harmonic oscillators interacting with a common reservoir is given in
\cite{Chou07}). In other words, the part of the squeezed state
relating to the relative variable is immune to the environment,
while that of the center-of-mass experiences severe dissipation and
noise from the environment. This results in the solution
(\ref{steady}). One can verify that Eq.~(\ref{steady}) is indeed a
decoherence-free squeezed state, i.e., $|\psi _{DFS}\rangle \propto
e^{x(a_{1}^{\dag }-a_{2}^{\dag })^{2}}|00\rangle $. If such a state
serves as the quantum channel, the quantum channel is free from the
vacuum fluctuation, and the entanglement is preserved.

\section{Numerical analysis of the non-Markovian entanglement dynamics}

The full entanglement dynamics of the squeezed-state quantum channel
is determined by the reduced density matrix obeying the master
equation (\ref{mas}). To solve the master equation, we must find
first the time-dependent coefficients contained in the master
equation, the shifted frequency $\Omega(t)$ and the shifted coherent
coupling $\Omega^{\prime }(t)$, as well as the individual and
correlated decay rates $\Gamma(t)$ and $\Gamma^{\prime }(t)$. These
coefficients are completely determined by the functions $u(t)$ and
$v(t)$ as the solutions of the dissipation-noise equations
(\ref{EOM}) via (\ref{soluen}). However, the dissipation-noise
equations have to be solved numerically for the general
environmental spectral density (\ref{spectral}).

\begin{figure}[tbp]
\begin{center}
\scalebox{1.05}{\includegraphics{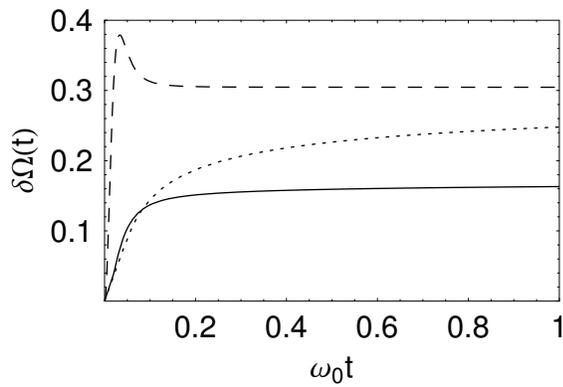}}
\end{center}
\caption{Time-dependence of the frequency shift $\protect\delta
\Omega (t) $ ($=\protect\delta \Omega ^{\prime }(t)$) induced by the
environment with spectral density (\protect\ref{spectral}) of $n=1$
for the Ohmic (solid line), $n=3$ for the super-Ohmic (dashed line),
and $n=1/2$ for the sub-Ohmic (dotted line). The asymptotical value
$\protect\delta \Omega (t \rightarrow \infty) = \protect\eta
\protect\omega_c, \protect\sqrt{\protect \pi}\protect\eta
\protect\omega_c$ and $2\protect\eta \protect\omega_c$ for the
Ohmic, the super-Ohmic, and the sub-Ohmic, respectively. The
parameters in (\protect\ref{spectral}) are taken as $\protect\eta
=0.005 $ and $\protect\omega _{c}=30.0 \omega _{0}$, while the
coupling constant between the cavity fields as $\protect\kappa
=0.5\omega _{0}$. } \label{shift}
\end{figure}

\begin{figure}[tbp]
\begin{center}
\scalebox{1.05}{\includegraphics{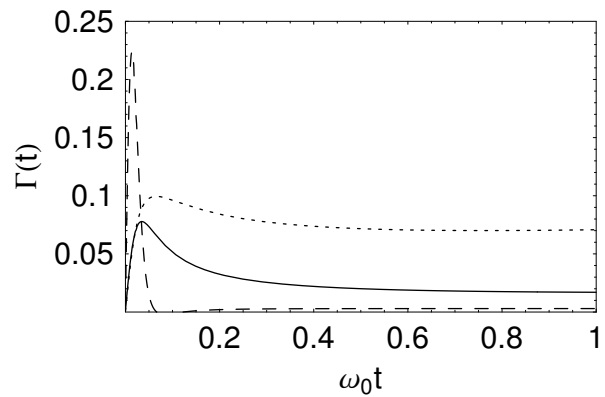}}
\end{center}
\caption{Time-dependence of the decay rate $\Gamma(t)$
($=\Gamma^{\prime }(t)$) with $n=1$ for the Ohmic (solid line),
$n=3$ for the super-Ohmic (dashed line), and $n=1/2$ for the
sub-Ohmic (dotted line) environments. The input parameters in the
numerical calculation are the same as in Fig.~1. } \label{decay}
\end{figure}

In Figs. \ref{shift} and \ref{decay}, we plot the numerical results
for the frequency shift $\delta \Omega (t)\equiv \omega _{0}-\Omega
(t)$ and decay rate $\Gamma (t)$ of the individual cavity field. We
choose three different spectral densities: $n=1, 1/2$, and 3 for the
Ohmic, the sub-Ohmic, and the super-Ohmic spectral densities,
respectively. Since the two cavity fields are considered to be
identical ($\omega_1=\omega_2=\omega_0$) and interact homogeneously
with the common environment ($g_{1k}=g_{2k}=g_k$), the
environment-induced shifts of the field frequencies and the coherent
coupling between the two cavity fields are equal, i.e., $\delta
\Omega (t)=\delta \Omega ^{\prime }(t)$ where $\delta \Omega
^{\prime }(t)\equiv \kappa -\Omega ^{\prime }(t)$. The individual
and correlated decay rates are also equal to each other, $\Gamma
(t)=\Gamma ^{\prime }(t)$,  for the same reason. From Figs.
\ref{shift} and \ref{decay}, we find that the dissipation-noise
dynamics is characterized by two time scales: $\tau _{1}=\omega
_{c}^{-1}$ (the shortest time scale of the environment) and $ \tau
_{2}=\omega _{0}^{-1}$ (the time scale of the cavity fields). When $
t<\tau _{1}$, both coefficients $\delta \Omega(t)$ and $\Gamma(t)$
grow very quickly. After $\tau _{1}$, they approach the
corresponding asymptotical values gradually in the time scale of
$\tau _{2}$. We should point out that the asymptotic values of
$\delta \Omega(t)$ and $\Gamma(t)$ in the Ohmic spectral density
reproduce the Markovian limit, as we have shown in our previous work
\cite{An07}). Compared with the super-Ohmic case, the sub-Ohmic
dissipation shows a slower asymptotical tendency in the time scale
$\tau _{2}$. This is because, in the super-Ohmic case, the
short-time correlation (the ultraviolet mode) is dominant, while in
the sub-Ohmic case, the long-time correlation (the infrared mode)
becomes important.

In fact, Fig. \ref{decay} also tells us that the decay rate grows
very fast in the time scale $\tau _{1}$ and develops a jolt
\cite{Hu92}. This peak manifests the significant effect of the
non-Markovian dynamics, especially for the super-Ohmic case. In the
usual Born-Markovian approximation used in the literature
\cite{Jakub04,An05,Rossi06,Adesso}, the back action of the
environment on the system is completely ignored by the assumption of
the response time of the environment being much smaller than the
characteristic time ($\tau _{2}$) of the system. The decay rate
becomes then time independent. The time dependence of the decay rate
in the exact master equation (\ref{mas}) contains the full memory
effect of the system interacting with the environment, as a result
of the non-Markovian dynamics. The non-Markovian entanglement
dynamics of the quantum channel thus becomes transparent due to the
presence of the time dependence of these coefficients in the time
scale $\tau _{1}$. This short-time correlation will influence
strongly the later-time entanglement dynamics of the squeezed state.

\begin{figure}[tbp]
\begin{center}
\scalebox{1.05}{\includegraphics{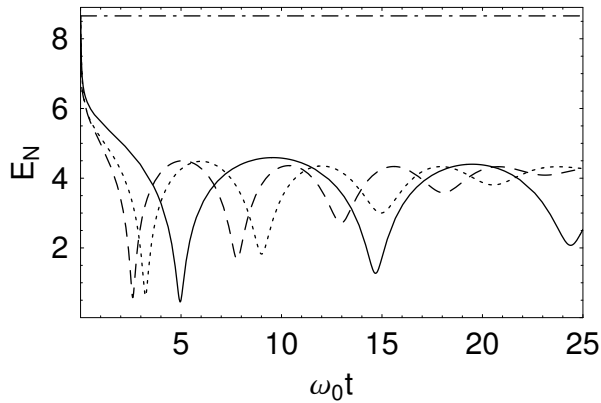}}
\end{center}
\caption{Time evolution of the logarithmic negativity $E_{N}(t)$
without the noise (dash-dotted line), and with the Ohmic (solid
line), the super-Ohmic (dashed line), and the sub-Ohmic (dotted
line) noise environments, in which the two cavity fields initially
have no coupling to each other ($\protect\kappa=0$). The other input
parameters are still the same, and the squeezing parameter $r=3.0$ }
\label{ent2}
\end{figure}

\begin{figure}[tbp]
\begin{center}
\scalebox{1.05}{\includegraphics{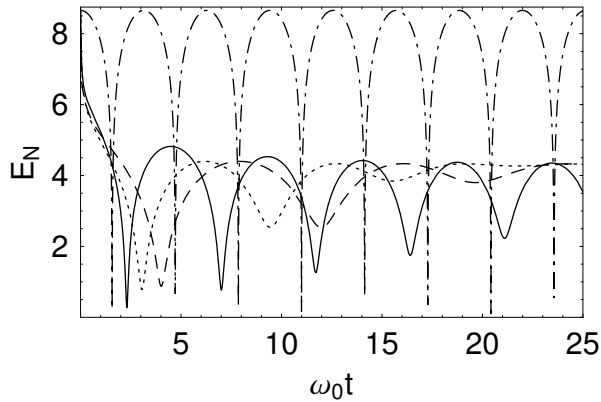}}
\end{center}
\caption{Time evolution of the logarithmic negativity $E_{N}(t)$
without the noise effect (dash-dotted line), and with the Ohmic
(solid line), super-Ohmic (dashed line), and sub-Ohmic (dotted line)
noise environments. The input parameters are the same as in Figs.~3,
except for $\kappa=0.5 \omega_{0}$ }\label{ent}
\end{figure}

In Figs.~\ref{ent2} and \ref{ent}, we plot the time evolution of the
logarithmic negativity for the entanglement dynamics of the squeezed
state. Figure \ref{ent2} describes the case where the two cavity
fields are initially decoupled ($\kappa=0$). As one would expect, in
the absence of vacuum fluctuations the entanglement does not change
in time. Figure \ref{ent} shows the coupling cavity fields
($\kappa=0.5$), where the entanglement undergoes a lossless periodic
oscillation. When the vacuum fluctuation is taken into account, the
entanglement dynamics is significantly changed as we see from
Figs.~\ref{ent2} and \ref{ent}. On one hand, the vacuum fluctuation
induces entanglement oscillations (see Fig.~\ref{ent2}) or shifts
the frequencies of the entanglement oscillations (see
Fig.~\ref{ent}) due to the effect of the shifted two-mode coupling
$\delta \Omega^{\prime }(t)$ (given in Fig.~\ref{shift}). We should
mention that the entanglement oscillations have also been found for
two uncoupled harmonic oscillators interacting with two independent
reservoirs \cite{Maniscalco07}. On the other hand, the amplitude of
the entanglement oscillation is suppressed gradually and tends to
its asymptotical value ($=r/\ln2$) arising from the joint
dissipation effects of the individual and correlated decay rates $
\Gamma(t)$ and $\Gamma^{\prime }(t)$.  Furthermore, the dissipation
will also erase the oscillation of entanglement arising from the
coherent coupling between the two cavity fields. These decoherence
effects are consistent with that obtained from the discrete qubit
models \cite{AnPA}. It shows again that the asymptotic value of the
entanglement reproduces the result in the Markovian limit
\cite{Jakub04}.

One may also see from Figs. \ref{ent2} and \ref{ent} that the
entanglement dynamics behaves differently for three different
spectral densities, as well as for the two coupled and uncoupled
cavity fields. Comparing with the oscillation behaviors presented in
Figs.~\ref{ent2} and \ref{ent}, it shows that the order of the
entanglement oscillations with three different spectral densities is
reversed for the coupled and uncoupled cavity fields. This behavior
comes mainly from the fact that the super-Ohmic environment induces
the strongest entanglement oscillation; next is the sub-Ohmic case;
while the Ohmic environment causes a relatively weak entanglement
oscillation, as shown in Fig.~\ref{shift}. By the definition
$\Omega'(t)\equiv \kappa -\delta \Omega'(t)$, it is easy to check
that, for the uncoupled cavity fields, $\kappa=0$ so that
$|\Omega'_{sup}(t)|> |\Omega'_{sub}(t)|>|\Omega'_{ohm}(t)|$. This
leads to the frequencies of the entanglement oscillations $f_{sup}>
f_{sub}>f_{ohm}$, as shown in Fig.~\ref{ent2}. While for the coupled
cavity fields, $\kappa=0.5$ so that
$|\Omega'_{sup}(t)|<|\Omega'_{sub}(t)|<|\Omega'_{ohm}(t)|$. This
gives the frequency ordering $f_{sup}< f_{sub}<f_{ohm}$, as plotted
in Fig.~\ref{ent}. This indicates that the non-Markovian induced
entanglement oscillations can be quite different for the two coupled
or uncoupled two entangled cavity fields.

Meanwhile, the entanglement oscillation is sustained for the longest
time in the Ohmic case, while the super- and sub-Ohmic environments
cause severe decoherence. The sub-Ohmic (low-frequency) vacuum
fluctuation induces a strong dissipative dynamics (the largest decay
rate, as in Fig. \ref{decay}), and therefore results in a fast
decoherenc of the entanglement of the squeezed state, as expected.
However, a remarkable result occurs in the super-Ohmic case where
the entanglement and its oscillation are also strongly suppressed.
In contrast to the sub-Ohmic case, the decay rate in the super-Ohmic
case is almost negligible except for a sharp peak in the short-time
scale $\tau_1$, as shown in Fig. \ref{decay}. We find that this
short-time sharp peak induces a significant contribution to the
entanglement decoherence in the quantum channel.  This decoherence
effect is a manifestation of the memory dynamics between the system
and the environment. It is this non-Markovian dynamics that causes a
rapid decoherence of the entanglement of the squeezed state in the
super-Ohmic environment. We should note that, within the time scale
$\tau_1=\omega_c^{-1}$, the initial peak of the decay rates also
exists in the Ohmic and sub-Ohmic cases although it is not as strong
as in the super-Ohmic case. This initial "jolt" in the decay rates
is indeed a general feature of the non-Markovian processes for the
decoherence enhancement, as pointed out first by Hu \textit{et al}.
\cite{Hu92} in the study of quantum Brownian motion. The results we
obtained in this work demonstrate that the non-Markovian dynamics
also speed the decoherence of the entanglement in
continuous-variable quantum channels.

\section{Summary}

In the present work, we have studied the detrimental effects of the
environment on the continuous-variable quantum channel in terms of
the entangled two-mode squeezed state. Using the Feynman-Vernon
influence functional theory in the coherent-state path integral
representation, we derive the exact master equation for the two
cavity fields under the influence of vacuum fluctuation \cite{An07}
and then investigate the non-Markovian entanglement dynamics of the
two-mode squeezed state quantum channel utilized in quantum
teleportation \cite{Furusawa98} for three different spectral
densities, the Ohmic, the sub-Ohmic, and the super-Ohmic
non-Markovian environments. Very recently, a similar exact master
equation has also been derived for two harmonic oscillators linearly
coupled to a thermal bath where the entanglement dynamics is studied
in the Markovian approximation \cite{Chou07}.

We numerically study the non-Markovian entanglement dynamics of the
quantum channel based on the exact master equation (\ref{mas}) for
three different noise environments. Our numerical result indicates
that the entanglement dynamics behaves different for the different
environmental spectral densities which leads to significant
distinctness in the time-dependent behavior of the dissipation-noise
function, in particular, within the short time scale $\tau_{1}$ of
the environment. For Ohmic environment the system shows the longest
quantum coherence because of the weak time-dependent dissipation of
the entanglement dynamics of the squeezed state. In the sub-Ohmic
case the squeezed state has a strong dissipation dynamics
(corresponding to a large decay rate) induced mainly by the
low-frequency noise of the environment, which results in fast
decoherence for the entanglement dynamics of the squeezed state. The
most significant evidence of the non-Markovian dynamics occurs in
the super-Ohmic environment in which the strong non-Markovian
process near the short time scale ($\tau_1$) speeds the decoherence
of the entanglement. These non-Markovian properties are indeed
consistent with the non-Markovian phenomena explored in quantum
Brownian motion \cite{Hu92}.

We may also point out that for the squeezed-state quantum channel
considered in this paper, both the asymptotical and the numerical
solutions show that one half of the initial entanglement carried by
the squeezed state will be retained regardless of the spectral
density of the environment. This is consistent with the solution in
the Markovian limit \cite{Jakub04}. This result depends only on the
structure of the initial squeezed state as well as the property of
the homogeneous coupling between the system and the environment.
Thus the asymptotical state (\ref{steady}) is indeed a
decoherence-free entangled squeezed state in our model, which may
serve as a noiseless quantum channel for further applications in
quantum communication. But it should be pointed out that, if the two
cavity fields couple with two independent reservoirs, the above
decoherence-free state no longer exists and the remaining
entanglement will eventually be lost completely \cite{Maniscalco07}.
As our concentration is on the optical cavity fields, we have
considered only the zero-temperature environment. A more general
case, e.g., with the environment at a finite temperature, could
hopefully be figured out by a similar approach to the derivation of
Eq.~(\ref{prord}). As robustness of the quantum channel is essential
in view of decoherence, we hope that our consideration of
non-Markovian entanglement dynamics in this paper provides useful
information for experimental designs of quantum-information
protocols.

\section*{Acknowledgement}

We would like to thank Professor B.~L. Hu, Dr. M.~T. Lee and W.~Y.
Tu for useful discussions. This work is supported by the National
Science Council of Taiwan under Contract No.~NSC-95-2112-M-006-001,
No.~NSC-94-2120-M-006-003, and No.~NSC-96-2119-M-006-001, and NNSF
of China under Grants No. 10604025 and Lzu05-02.

\section*{Appendix: the coefficients of $\protect\rho (\boldsymbol{\bar{
\protect\alpha}}_{f},\boldsymbol{\protect\alpha }_{f}^{\prime };t)$ and
their asymptotical behaviors}

The explicit form of $\rho (\boldsymbol{\bar{\alpha}}_{f},\boldsymbol{\alpha
}_{f}^{\prime };t)$ is obtained from Eq. (\ref{rout}) by the evaluation of
the integration. The final solution is given by Eq. (\ref{final}) with the
parameters
\begin{widetext}
\begin{eqnarray*}
b_{0} &=&\frac{1}{\cosh ^{2}r\sqrt{1-2\tanh ^{2}r(m^{2}+n^{2})+\tanh
^{4}r(m^{2}-n^{2})^{2}}}, \\
b_{1} &=&\frac{e\tanh ^{4}r[(un+vm)^{2}+(um+wn)^{2}]+c\tanh
^{3}r(un+vm)(um+vn)}{1-2\tanh ^{2}r(m^{2}+n^{2})+\tanh
^{4}r(m^{2}-n^{2})^{2}
}+\tanh ruv, \\
b_{2} &=&\frac{e\tanh ^{2}r(\bar{u}^{2}+\bar{v}^{2})+c\tanh
r\bar{u}\bar{v}}{
1-2\tanh ^{2}r(m^{2}+n^{2})+\tanh ^{4}r(m^{2}-n^{2})^{2}}, \\
b_{3} &=&\frac{e[-4\tanh ^{4}r(un+vm)(um+vn)]+c\{-\tanh
^{3}r[(un+vm)^{2}+(um+vn)^{2}]\}}{1-2\tanh ^{2}r(m^{2}+n^{2})+\tanh
^{4}r(m^{2}-n^{2})^{2}}-\tanh r(u^{2}+v^{2}), \\
b_{4} &=&\frac{e\{-2\tanh ^{3}r[\bar{u}(un+vm)+\bar{v}(um+vn)]\}+c\{-\tanh
^{2}r[\bar{u}(um+vn)+\bar{v}(un+vm)]\}}{1-2\tanh ^{2}r(m^{2}+n^{2})+\tanh
^{4}r(m^{2}-n^{2})^{2}}, \\
b_{5} &=&\frac{e\{2\tanh ^{3}r[\bar{u}(um+vn)+\bar{v}(un+vm)]\}+c\{\tanh
^{2}r[\bar{u}(un+vm)+\bar{v}(um+vn)]\}}{1-2\tanh ^{2}r(m^{2}+n^{2})+\tanh
^{4}r(m^{2}-n^{2})^{2}}, \\
b_{6} &=&\frac{e(-4\tanh ^{2}r\bar{u}\bar{v})+c[-\tanh
r(\bar{u}^{2}+\bar{v} ^{2})]}{1-2\tanh ^{2}r(m^{2}+n^{2})+\tanh
^{4}r(m^{2}-n^{2})^{2}},
\end{eqnarray*}
\end{widetext}
where $c=1-\tanh ^{2}r(m^{2}+n^{2})$, $e=\tanh rmn$,
$m=\bar{u}u+\bar{v}v-1$, and $n=\bar{u}v+\bar{v}u$. In the long
time limit, $u(t\rightarrow \infty )=v(t\rightarrow \infty
)=\frac{e^{-i(\omega _{0}-\kappa )t }}{2}$. Then
\begin{eqnarray*}
b_{0}(t\rightarrow\infty ) &=&\frac{1}{\cosh r}, ~ \text{\ \
}b_{4}(t\rightarrow\infty )= b_{5}(t\rightarrow\infty )=0, \\
\text{\ \ }b_{1}(t\rightarrow\infty ) &=&b_{2}^{\ast
}(t\rightarrow\infty ) = \frac{-b_{3}(t\rightarrow\infty )}{2} \\
&=&\frac{-b_{6}^{\ast }(t\rightarrow\infty )}{2}=\frac{\tanh
re^{-2i(\omega_{0}-\kappa )t}}{4}.
\end{eqnarray*}

\end{document}